%% file: main.tex
\documentclass{article}

\usepackage{arxiv}

\usepackage[utf8]{inputenc} 
\usepackage[T1]{fontenc}    
\usepackage{hyperref}       
\usepackage{booktabs}
\usepackage{float}
\usepackage{graphicx}
\usepackage[dvipsnames]{xcolor}
\usepackage{tikz}
\usepackage[caption=false,font=normalsize,labelfont=sf,textfont=sf]{subfig}
\usepackage[numbers,sort&compress]{natbib}
\usepackage{psfrag}
\usepackage{array}
\usepackage{chngpage}
\usepackage{indentfirst}
\usepackage{amsmath}
\usepackage{amssymb}
\usepackage{epstopdf}
\usepackage{enumitem}
\usepackage{multirow}
\usepackage{siunitx}
\usepackage{pgfplots}
\usepackage{pgfplotstable}

\pgfplotsset{compat=1.14}
\usepgfplotslibrary{dateplot}
\definecolor{darkerBlue}{RGB}{0,102,204}
\definecolor{brewer_div_gold}{RGB}{216,179,101}
\definecolor{brewer_div_green}{RGB}{90,180,172}
\definecolor{PuBu-5-1}{RGB}{241,238,246}
\definecolor{PuBu-5-2}{RGB}{189,201,225}
\definecolor{PuBu-5-3}{RGB}{116,169,207}
\definecolor{PuBu-5-4}{RGB}{43,140,190}
\definecolor{PuBu-5-5}{RGB}{4,90,141}
\definecolor{Greys-3-3}{RGB}{99,99,99}
\definecolor{PuRd-5-1}{RGB}{241,238,246}
\definecolor{PuRd-5-2}{RGB}{215,181,216}
\definecolor{PuRd-5-3}{RGB}{223,101,176}
\definecolor{PuRd-5-4}{RGB}{221,28,119}
\definecolor{PuRd-5-5}{RGB}{152,0,67}
\definecolor{Reds-7-1}{RGB}{254,229,217}
\definecolor{Reds-7-2}{RGB}{252,187,161}
\definecolor{Reds-7-3}{RGB}{252,146,114}
\definecolor{Reds-7-4}{RGB}{251,106,74}
\definecolor{Reds-7-5}{RGB}{239,59,44}
\definecolor{Reds-7-6}{RGB}{203,24,29}
\definecolor{Reds-7-7}{RGB}{153,0,13}
\definecolor{Blues-7-1}{RGB}{239,243,255}
\definecolor{Blues-7-2}{RGB}{198,219,239}
\definecolor{Blues-7-3}{RGB}{158,202,225}
\definecolor{Blues-7-4}{RGB}{107,174,214}
\definecolor{Blues-7-5}{RGB}{66,146,198}
\definecolor{Blues-7-6}{RGB}{33,113,181}
\definecolor{Blues-7-7}{RGB}{8,69,148}

\newcommand{\kwh}{kWh}

\title{Understanding Power and Energy Utilization in Large Scale Production Physics Simulations Codes}

\author{ \\Adam Bertsch $^1$,
  Michael R. Collette$^1$,
  Shawn A. Dawson$^1$,
  Si D. Hammond$^2$,
  Ian Karlin$^3$,\\
  M. Scott McKinley$^1$,
  Kevin Pedretti$^4$,
  Robert N. Rieben$^1$,
  Brian S. Ryujin$^1$,
  Arturo Vargas$^1${*},
  Kenneth Weiss$^1$\\
  \\
  Lawrence Livermore National Laboratory, Livermore, CA, USA$^1$, 
  \\National Nuclear Security Administration US Department of Energy, Washington, DC, USA$^2$
  \\NVIDIA, Santa Clara, CA, USA$^3$
  \\Sandia National Laboratory, Albuquerque, NM, USA$^4$\\
}

\renewcommand{\headeright}{}
\renewcommand{\undertitle}{Technical Report}

\hypersetup{
pdftitle={Understanding Power and Energy Utilization in Large Scale Production Physics Simulations Codes},
}

\begin{document}
\maketitle

\begin{abstract}
Power is an often-cited reason for the move to advanced architectures on the path to Exascale computing. This is due to practical considerations related to delivering enough power to successfully site and operate these machines, as well as concerns about energy usage while running large simulations. Since obtaining accurate power measurements can be challenging, it may be tempting to use the processor thermal design power (TDP) as a surrogate due to its simplicity and availability. However, TDP is not indicative of typical power usage while running simulations. Using commodity and advanced technology systems at Lawrence Livermore and Sandia National Labs, we performed a series of experiments to measure power and energy usage in running simulation codes. These experiments indicate that large scale Lawrence Livermore simulation codes are significantly more efficient than a simple processor TDP model might suggest.

This work performed under the auspices of the U.S. Department of Energy by Lawrence Livermore National Laboratory un\
der Contract DE-AC52-07NA27344.  
\end{abstract}

\keywords{HPC, \and energy \and power}

\section{Introduction}
Power requirements and energy usage are important factors in the siting, operating costs and environmental impact of a supercomputer.
Energy, which we measure in joules, is the ability to create a change within the circuit system, while power, measured in watts, is the rate at which energy is consumed. Given a power rate, energy is derived as the integral of power over time. Exascale machines require tens of megawatts to operate and the facilities hosting them are undergoing costly renovations to accommodate their power needs.
Given these high costs, it is important to understand the relative power consumption of various processor options for different workloads.

One approach to lower power usage is through the use of accelerators.  Many TOP500 systems today use GPUs for both performance benefits and lower power consumption.
For FLOP-heavy workloads that are similar to the LinPACK benchmark, they provide clear performance per watt advantages as is evidenced by the top machines on recent~\cite{top500green} list.

Studies looking at real applications abound and comparisons between various systems are frequent. Most only focus on the processor component of power and energy while some do look at full system power. In addition, some look at the impact of code optimization on power and energy usage at the
processor level. In this paper we investigate the following areas not covered in previous studies:

\begin{itemize}
  \item Holistic power measurements that include switches, power supply losses and processors that show where all the power goes in running HPC applications.
  \item The relationship between TDP and measured power usage.  We show that system TDP often means significantly more power is provisioned compared to what is needed to run the system in production.
  \item Cross platform energy breakdowns and comparisons and energy/performance tradeoffs.
\end{itemize}
In addition, we cover some old ground with new large production simulation codes as we look at the effects of code optimization on power and energy.

Our overall results show that processor TDP is a poor proxy for practical system power/energy usage. Other components may dominate usage, and processor power is often significantly less than the stated peak for many real HPC workloads.

{
The studies in the article consider the power and energy usage of three production simulation codes across several computing platforms. 
Given the variations in test problems for each code and the methods available for measuring system power consumption, we organize the remainder of this article as follows:
  We begin by exploring the use of processor TDP as a surrogate for understanding application power usage, and review prior studies on application energy usage.
  Next, we present an overview of the computing platforms in our study and describe our methodology for measuring power consumption on these systems.
  This is followed by a detailed discussion of the production simulation codes and test problems employed in our experiments.
  Finally, we examine the interplay between performance and optimizations on GPU-based platforms and establish metrics for identifying energy breakeven points in a cross platform comparison.
}

\subsection{Limitations of processor thermal design power as a surrogate}

\emph{Thermal design power (TDP)} is the maximum amount of heat that a component is
designed to generate. It is included in the standard specification for many
computer components, including processors. Due to simplicity and accessibility, it may be tempting to use a
processor's TDP as a surrogate for power usage in cross platform evaluations,
where the node TDP is not readily available. Using this
single number can make it easy to make performance targets for energy breakeven,
i.e.\ if one processor's TDP is two times higher than another's, the second processor
merely needs to run two times faster to be equally energy efficient.
Such a simplified analysis would ignore differences in overall node design or
assume that the whole node's TDP is dominated by the processor, which is
not necessarily true. It should also be noted that the TOP500 list does not
list TDP of any machine, but instead the actual power used when running the
LinPACK benchmark.

\subsection{Related work on HPC power and energy studies}
As an example of using TDP as a power surrogate, the work of~\cite{franko2015cfd} performs a cross platform study using a computational fluid dynamics code. They compare energy consumed to complete a given simulation across four different platforms. In the article, consumed energy is derived by multiplying processor TDP by execution run time which would imply the simulation is running at power capacity of the card. In the work we present here, we find that this is not typically the case. A similar assumption is made in the work of~\cite{lannelongue2021green} which aims to develop a framework for quantifying carbon footprint from computations.


Additional studies on power energy utilization include the work of~\cite{price2016optimizing} which investigates the performance per watt on GPUs.  They specifically catalog the effect of temperature and supply voltage showing that performance per watt can be increased by 37--48\% over default settings by lowering supply voltage and increasing clock frequency while maintaining low die temperatures.

To assist in understanding energy usage at the processor level, the MSR-Safe library by ~\cite{mcfadden2019msr} and Variorium library by ~\cite{Variorum} offer interfaces for the Intel and AMD processors.

\cite{enos2010quantifying} develop a hardware methodology for measuring and comparing performance per watt for specific applications for both CPU and GPU implementations.  They show that for the same workload, performance per watt can be improved by running the same application after porting to the GPU.

\cite{patel2020does} explore the power characteristics of typical HPC jobs during the approach to the Exascale era.  They show that as HPC systems become increasingly power constrained, a data-driven approach to HPC application power characteristics can be used to make more effective use of HPC systems.

\cite{kamil2008power} show that the high performance Linpack (HPL) benchmark is a useful proxy for compute intensive kernels in multiple HPC workloads for the purpose of predicting power consumption, while it is not a useful proxy for projecting application performance.  They describe the increasing need to establish practical methods for measuring application power usage in-situ to understand behavior in a post Dennard scaling era.

We extend these works with methods for extracting power and energy data on some production applications of interest -- our large simulation codes -- which can be collected during normal production runs on the Sierra supercomputer.  We show that the energy to solution advantage for GPU-enabled applications is significant over CPU-only solutions. We also show that GPU TDP is not a valid proxy for power usage on these production applications due to the significant and variable differences between GPU TDP and GPU average power for these production applications. 
{Furthermore, the methodology presented in this work served as a foundation for the work of}~\cite{horwitz2024estimating}~{where the author recognizes the importance of actual power measurements to quantify carbon footprint for computational fluid dynamics (CFD) simulations.}

\section{Overview of computing platforms}
\begin{figure*}[t]
  \centering
  \includegraphics[width=0.65\textwidth]{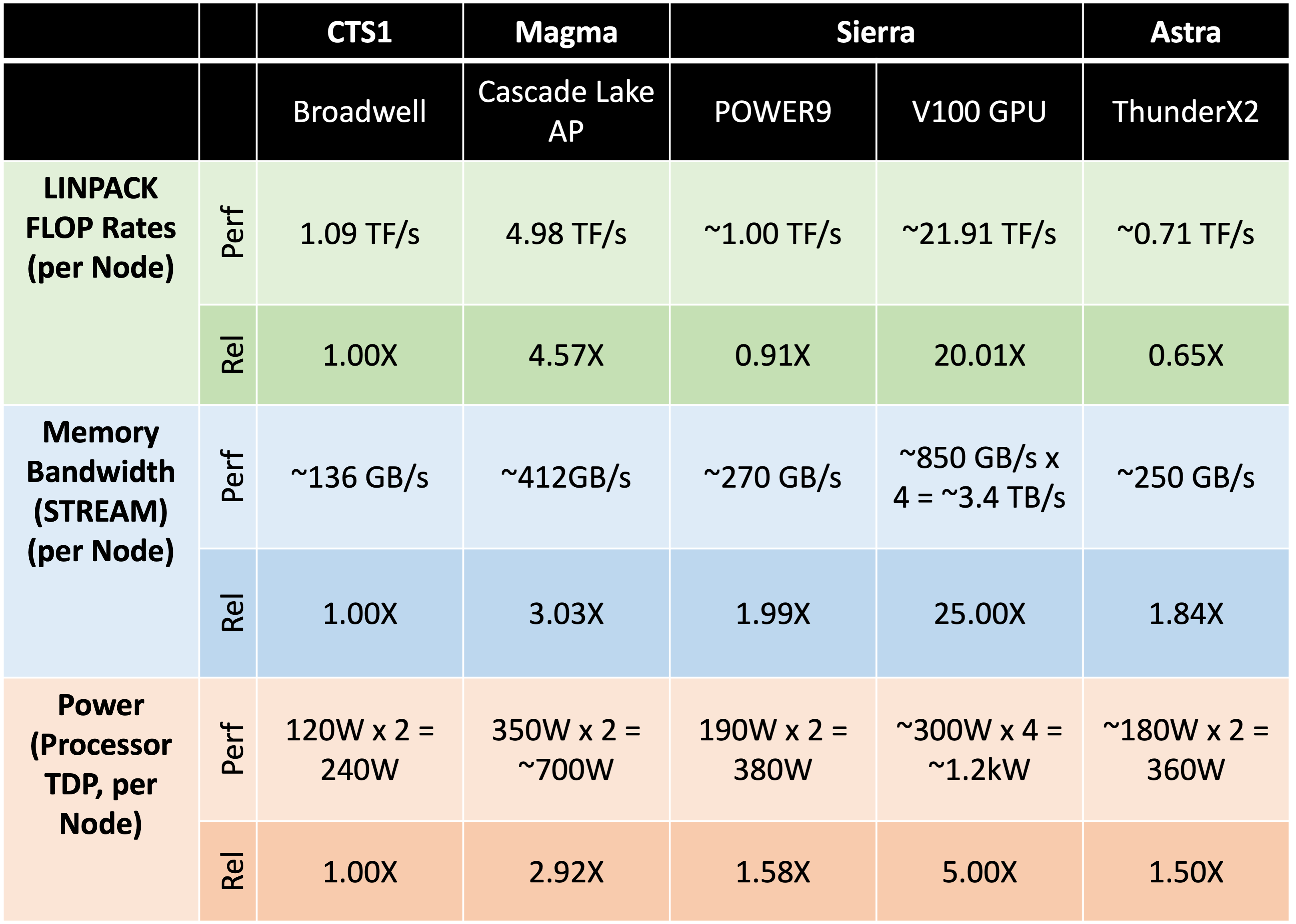}
  \caption{Computing platforms used in this work. The commodity based platform (CTS-1) serves as our baseline for comparisons. {Due to limitations in measuring power and energy we are unable to provide a per rack power  or a per node TDP.}  }
  \label{fig:PlatformComaparison}
\end{figure*}

In this work, we study power and energy usage on three computing platforms from Lawrence Livermore National Lab -- a commodity technology system (CTS-1) consisting of Intel Xeon E5-2695 v4 2.1GHz (Broadwell) CPUs, Magma consisting of Intel Xeon Platinum 9242 48C 2.3GHz (Cascade Lake AP) CPUs, and Sierra IBM POWER9 22C 3.1GHz and NVIDIA Volta GV100 GPUs, as well as a fourth platform from Sandia National Laboratory -- the ARM-based Astra cluster consisting of Marvell ThunderX2 CN9975-2000 28C 2GHz (ThunderX2) CPUs. Figure~\ref{fig:PlatformComaparison} provides a comparison of reported LinPACK FLOP rates, memory bandwidth performance with a stream benchmark, and processor TDP per node. Throughout the table, the commodity platform (CTS-1) serves as our baseline for comparisons. Notably the CTS-1 processors have the smallest TDP, while graphics processing units have the highest. The LinPACK power usage is reported for each machine on the~\cite{top500} website. Memory bandwidth numbers were estimated using microbenchmarks and the processor TDP came from vendor specifications.

\subsection{Measuring power on Sierra}
The Sierra system has multiple touchpoints for the measurement of both power and energy.  Each 360-node section of the system is equipped with a wall plate power meter with 1 second time resolution.  The system also has node-level power measurement capabilities provided by the IBM Witherspoon compute node.  This node-level power measurement is integrated over the course of a compute job and recorded by the IBM Cluster System Management software and stored as a job energy value in a database.  The scale of Sierra meant that 360 node runs would be too large for some of the measurements in this work.  As a result, we installed an additional wall plate power management system on the switch components of one rack and over the course of large test runs we were able to determine the typical switch power usage to be 15 watts per node and the power supply and other losses to be 15 percent when comparing power data computed from CSM database energy values to the 360 node wall plate values.  The collected power data does not attempt to account for a percentage of the top-level switch power.

\subsection{Measuring power on CTS-1 and Magma}
Our CTS-1 and Magma systems use rack-level power rectifiers and DC power distribution within the rack.  As a result, power for this system may be measured at the rack level.
We performed dedicated runs using whole compute racks from the machine and collected power data with the help of system staff. The data from these power rectifiers includes swhich power and power supply losses within the rack but does not attempt to account for a percentage of the top-level switch power, allowing for a direct comparison against our other systems.  The rectifier power data was collected every minute through the runs in order to determine an energy value for the run.

\subsection{Measuring power on Astra}


The Astra supercomputer provides power and energy measurement capabilities at the processor, node, chassis, rack, and full system levels~\citep{grant2023enabling}. At the processor level, the Marvell ThunderX2 ARM processors used in the system provide extensive on-die voltage, power, frequency, and temperature measurements on a per-core basis. These measurements can be accessed in-band by users by instrumenting their code using the PowerAPI~\citep{grant2016standardizing} or by using vendor supplied tools. At the node and chassis levels, the HPE Apollo 70 server architecture provides out-of-band interfaces for measuring per-node power usage and a range of environmental sensors. At the rack level, the power distribution unit (PDU) in each rack provides a convenient location for measuring the total energy consumed by all the equipment in the rack, including the compute nodes, network switches, and other ancillary components. Lastly, at the full system-level, the 480VAC overhead bus bars and 208VAC PDUs that together supply power to the system include measurement capabilities that can be used to calculate the system's total power and energy usage.

\subsubsection{Tradeoffs of the different measurement points}

Each of the available power measurement points has different accuracy, precision, sampling frequency, and user interface characteristics that must be carefully considered when designing experiments and interpreting results. For example, the system's health monitoring infrastructure collects node-level power measurements via out-of-band interfaces that do not affect application performance, however the measurements obtained are low precision and low fidelity (e.g., quantized to multiples of 8 watts with 1/min sampling). This is appropriate for system monitoring activities, but it may not be appropriate for performing detailed application power usage experiments. In contrast, the rack-level PDUs provide billing grade energy measurements based on high internal sampling rates and $\pm 1\%$ overall accuracy. This provides high-quality aggregate rack-level measurements, but it is not possible to resolve the energy used by individual compute nodes.

\subsubsection{Experimental method presented in this paper}

The Astra results presented in this paper were gathered using rack-level energy measurements since this was the most closely comparable result to the experiments performed on the other systems. Each of the workloads was configured to run on either one or two full racks (72 or 144 compute nodes) and to execute for several hours of runtime. The jobs were run on a dedicated reservation of two compute racks that was pre-screened to ensure that all nodes were available and operating correctly. As each job ran, the job ID was noted so that the jobs start time, end time, and node list could be looked up from the batch scheduler logs. This information was used to retrieve the corresponding rack-level energy measurements from the system monitoring database, resulting in a series of 1/min timestamped energy measurements covering the jobs entire execution window. The low sampling rate does not induce significant error due to the long runtimes (e.g., a 3-hour job with 1/min energy sampling results in \textless 1\% error). The jobs total energy consumption can be calculated by subtracting the first measurement from the last, resulting in the total Joules consumed, or a power vs. time plot can be generated by examining the energy and timestamp deltas between adjacent measurements.

\section{Overview of tested simulation codes}
For many large scale LLNL simulation codes, running efficiently on Sierra required major code refactoring such as porting computational kernels and revisiting memory management strategies.
Abstraction layers such as RAJA~\citep{beckingsale2019raja} and memory resource managers such as Umpire~\citep{beckingsale2019umpire} have helped simplify the porting process with a single source code for different computing platforms, but still require developer expertise to ensure correctness and performance.
This has helped many of the LLNL codes which have successfully ported to GPUs realize major speedups compared to existing CPU based computing platforms~\citep{beckingsale2019raja}.
To run on GPUs, \emph{Ares} and \emph{Marbl} use RAJA and Umpire, while \emph{Imp} uses Umpire and a custom portability layer to enable the same capabilities as RAJA. As Ares has established GPU capabilities, we present experiments with various optimizations. As the refactoring of Marbl to GPUs has been a more recent effort, we present data capturing power usage as kernels were incrementally ported and optimized. Lastly, we perform a cross platform study using all three codes.  In our experiments all three codes were executed with MPI-only parallelism on CPU based platforms. Here we map one MPI per core processor core {using the standard distributed memory model}.  For the GPU platforms, we employed 1 MPI rank per GPU and used the CUDA backend of our abstraction layer for offloading to the device.

\subsection{Ares}
Ares is a massively parallel, multi-dimensional, multi-physics simulation code~\citep{bender_ares_2021}.
Its capabilities include ALE-AMR hydrodynamics,
radiation diffusion and transport, 3T plasma physics and high explosive
modeling. It has been used to model many different types of experiments,
such as inertial confinement fusion (ICF), pulsed power and high explosives.

For the cross-platform comparison study, Ares used a 3D multi-material ALE
hydrodynamics problem that contained 103 million zones and ran for 25,000 cycles.

For the optimization study, Ares used an ALE hydrodynamics problem that modeled
a Rayleigh-Taylor mixing layer in a convergent geometry. It was a 4$\pi$, 3D
simulation which contained 23.8 million zones and ran for 10,380 cycles.

\subsection{Marbl}
Marbl is a newer multi-physics simulation code at Lawrence Livermore National Lab. Some of its key capabilities include multi-material radiation hydrodynamics in both an Eulerian and an Arbitrary Lagrangian-Eulerian (ALE) framework~\citep{anderson2020multiphysics}. Marbl builds on modular physics and computer science packages such as Axom~\citep{Axom}, MFEM~\citep{mfem}, RAJA~\citep{beckingsale2019raja} and Umpire~\citep{beckingsale2019umpire} to achieve cross-platform performance~\citep{Vargas2022_ijhpca,Stitt2024_jfe}. A distinct feature of this code is its design choice of employing high-order numerical methods. In this study we exercise the high-order finite element multi-material ALE package to perform our power and energy studies.

As a test problem for a cross platform comparison, we choose the three-dimensional multi-material Triple-Point problem. For the CPU based platforms (CTS-1, Magma, and Astra), the problem was configured with a mesh consisting of 462 million quadrature points and was executed for 500 cycles. For the GPU based platform, Sierra, the problem was scaled to a mesh with 3.7 billion quadrature points and 5,000 cycles. The discrepancy in problem size stemmed from the large run-time differences between the platforms and the requirement of running the code long enough in order to measure power and energy usage.

Additionally, we were able to align this study with Marbl's GPU modernization effort, which enabled us to track the effects of incrementally offloading and optimizing kernels on power and energy usage. For this study, we exercised a three-dimensional shaped charge problem on a node of Sierra (4 NVIDIA V100's) for 1,000 cycles.

\subsection{Imp}

Imp is a new implicit Monte Carlo (IMC) thermal photon transport simulation code~\citep{Brantley2019}
which implements the standard IMC algorithm for time- and frequency-dependent x-ray photon transport as defined by Fleck and Cummings~\citep{fleck1971implicit}.
Some general features of Imp include photon sources, effective photon scattering, thermal photon emission, photon opacities, and source tilting.
Imp supports multiple mesh geometries and implements multiple parallel algorithms including MPI, OpenMP, and GPU parallelism.

The test problem used for the energy study is a half-hohlraum, a simplified 2D hohlraum simulation
modeling photon transport in a geometry motivated by laser-driven radiation-hydrodynamics
experiments.  This is further defined in~\cite{yee2021new}.

\section{Cross platform energy and power analysis}
To better understand how power and energy usage varies across the different platforms, we consider several approaches. One approach is to consider energy usage with respect to speedups, and the second is to compare throughput. Using the Ares code, we performed a strong scaling study on a multi-material ALE hydrodynamics problem, while Marbl and Imp examined energy required per a unit of work for a multi-material ALE hydro problem.
Since LinPACK corresponds to exceptionally heavy computational workloads, we believe the idle power and LinPACK can serve as lower and upper respectively bounds for scientific simulation codes.
Table~\ref{tab:idle_and_LinPACK_energy} compares processor TDP, idle power, and LinPACK power usage.


\input{tables/power_measurements}

Our studies measured energy in terms of joules, and we present it in terms of Kilowatt-hours, where
1~\kwh~=~$3.6~\cdot~10^6$~J.
Conversion to watts per node is given by
\[
\textbf{watts per node} = \frac{\text{joules}}{\text{seconds} \times \text{nodes}}.
\]
{The simulation runs included minimal I/O and ran enough cycles that the variation in compute behavior due to initialization and finalization of the problems should be minimal. Additionally, the behavior is pretty similar between cycles and thus would exhibit minimal variation}.  In this work we are interested in quantifying a required speedup in order to reach an {energy breakeven point}. Since we are taking actual energy measurements, we define the power ratio as:
\[
\textbf{PowerRatio}_{\text{Energy breakeven}} = \frac{\text{Measured Power on Sys 1}}{\text{Measured Power on Sys 2}}.
\]
The power ratio informs us of the required speedup needed between platforms to reach an energy breakeven point. {While the exact value of this metric will be application and problem dependent, it can still be a useful tool to identify trends between systems.}



\subsection{Ares}

\begin{figure*}[tbp]
  \centering
  \includegraphics[width=.75\linewidth]{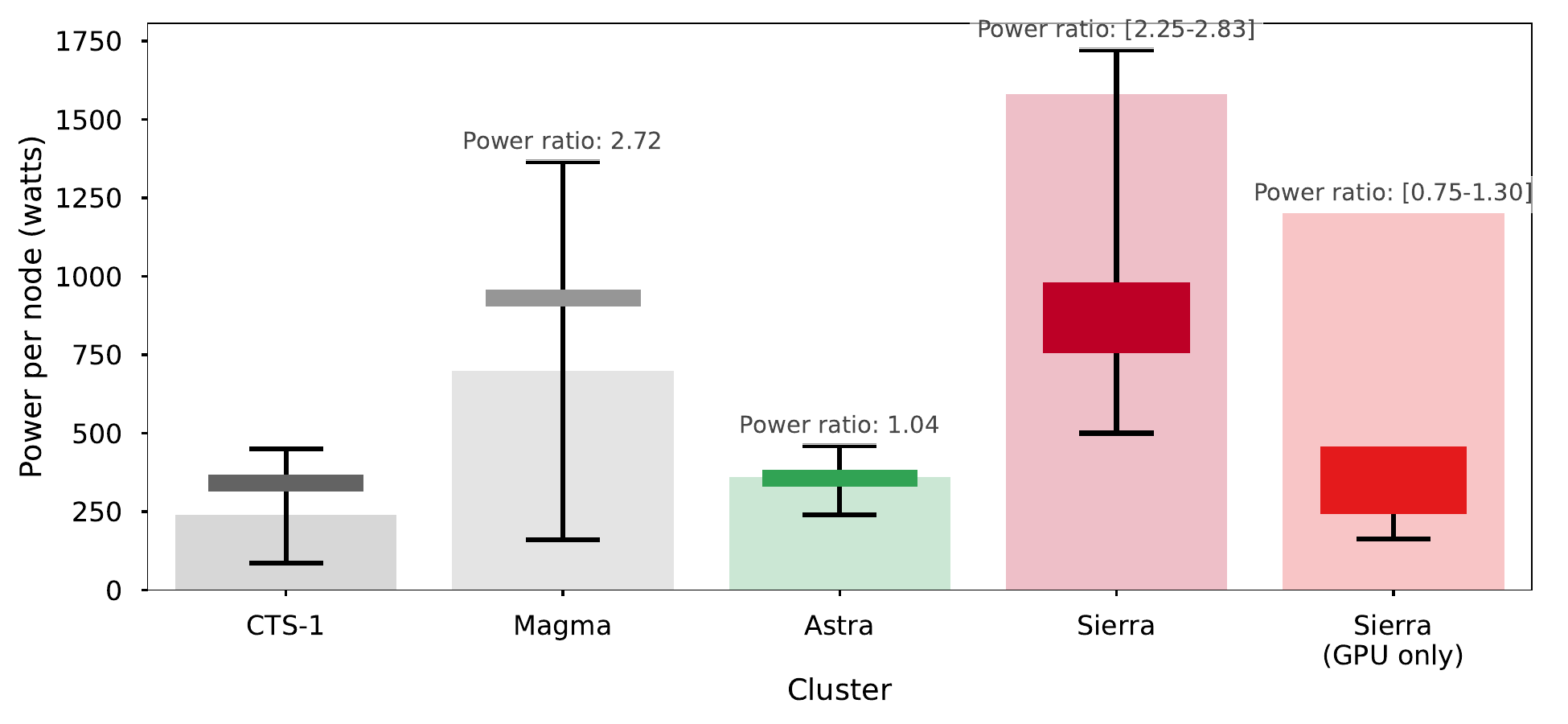}
  \caption{Ares power usage (dark bars) relative to machine idle (lower whisker) and LinPACK power (upper whisker) rates. {Bar thickness indicates the ranges of results when there are multiple runs on the platform.} Power ratio refers to the necessary speed up to break even in terms of energy usage compared to the CTS-1 machine. Light colored bars corresponds to processor TDP for a given platform (excluding rack-level measurements). LinPACK power utilization is observed to go beyond processor TDP as it consist of a more complete power measurement.}
  \label{fig:Measured power from Ares}
\end{figure*}

\input{tables/ares_energy}

We performed a strong scaling study of a multi-material ALE hydrodynamics problem
across all platforms studied, which was constrained by node counts needed to
get accurate power and energy data and the results are in Table~\ref{tab:ares_cross_platform_energy}.
GPU only power was derived using the IBM system monitoring tool and only computed for the ARES code
for the cross platform studies.

The range of power usage across all runs are summarized in Figure~\ref{fig:Measured power from Ares}.
The power usage on each platform for this
problem remains in a narrow band, relative to the spread of idle power and
LinPACK power {measured across all components of the node}. 
It is also apparent that the processor TDP is not generally reflective
of actual usage. The only clear commonality that can be seen in this data is
that the CTS-1 and Magma platforms have similar TDP to actual usage ratios.
These both contain Intel processors and have a similar node design, so that may
not be unexpected.

On every platform with multiple runs, the runs with the fewest number of nodes
is consistently the most energy efficient and offer the highest throughput of
work. Ares does not strong scale
perfectly, so, as the resources increase, the time does not decrease
proportionally. Although there is also a reduction in power per node as the code
is strong scaled, it does not reduce enough to offset the increased number of
nodes used.

One common metric for comparing platforms is to use a node-to-node comparison.
Due to limitations in the energy measuring methodology, there aren't exact
node count matches across platforms. For comparing Sierra to Astra, the closest
available is Sierra's 80 node run with Astra's 72 node run. Comparing those two
data points shows that Sierra has a 3.3x speedup over Astra and that it is
1.4x more energy efficient. The ratio of node power at these points is 2.15,
which suggests that the speedup needed to reach breakeven on a GPU for this
problem is only 2.15. It should also be noted that at this point, the Sierra
runs are strong scaled and less energy efficient than its other runs.

For the same metric on the CPU platforms, Astra's 72 node run and CTS-1's 62
node run are the closest. For those runs, there is a 1.3x speedup on Astra.
Astra is 1.13x more energy efficient than CTS-1. The ratio of the power between the
two runs is 1.04, so the gain in efficiency is almost entirely from the
faster runtime on Astra.

Another way to compare platforms is to look at equivalent node counts to get
the same answer in the same amount of time. Using this lens, the runs with the
closest duration are Sierra's 5 node run with Astra's 72 node run and Sierra's
10 node run with Astra's 144 node run. In these cases, Sierra is about 5.5x
more energy efficient than Astra for running the same problem in the same
amount of time. The ratio of node power between these runs is between 2.6 and
2.7, which is more than offset by the 14x difference in nodes being used.

\subsection{Marbl}
Marbl's cross platform study aimed to compare energy and power usage across the smallest number of nodes necessary for the highest fidelity energy and power estimates.
For the CPU based platforms, CTS-1, Magma, and Astra, the node counts were 62, 48, and 72 respectively. 
Prior to understanding the correction factor for Sierra ({as discussed earlier in the section on measuring power on Sierra}),
achieving the highest fidelity power measurements on Sierra required running on 360 nodes.

\begin{figure*}[tbp]
  \centering
  \includegraphics[width=.75\linewidth]{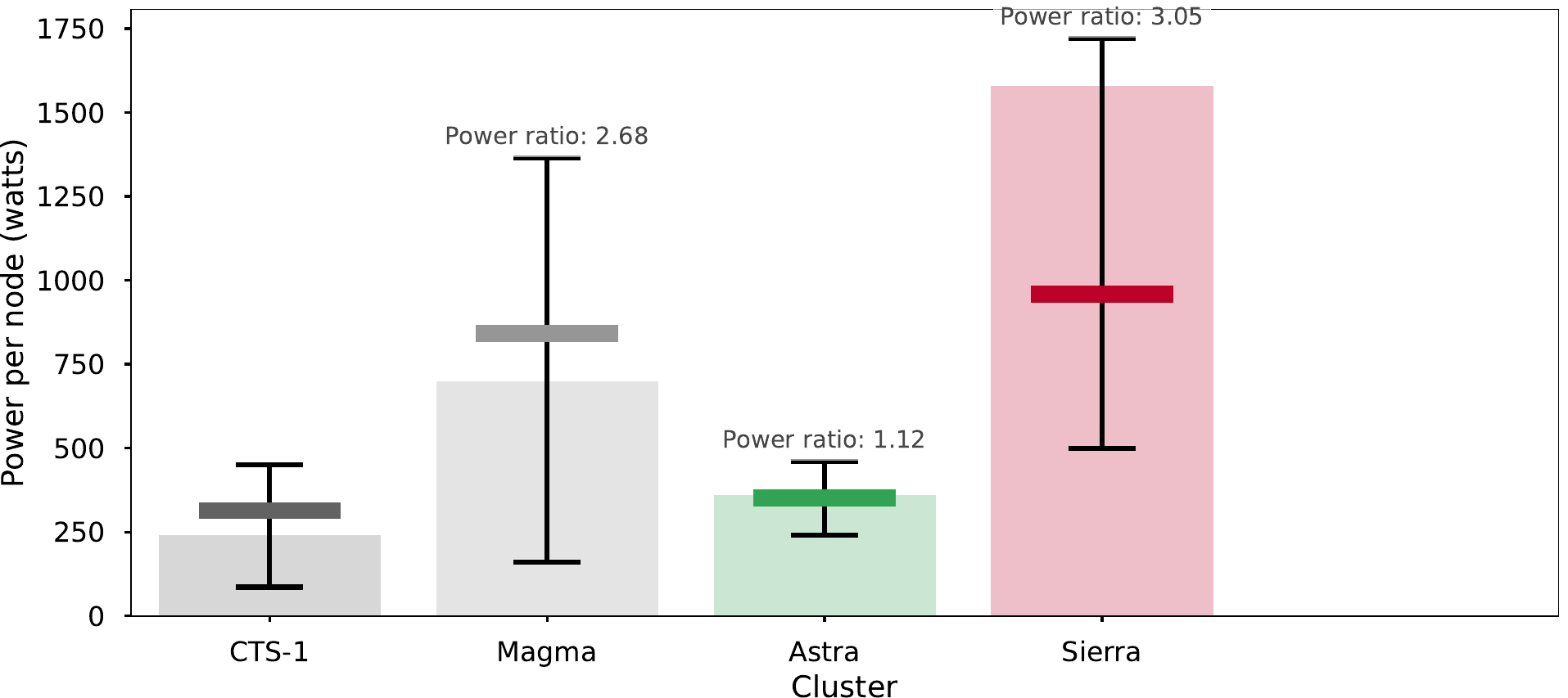}
  \caption{Marbl power (dark bar) requirements relative to machine idle (lower whisker) and LinPACK power (upper whisker) rates. Power ratio refers to the necessary speed up to break even in terms of energy usage compared to the CTS-1 machine. Light colored bars corresponds to processor TDP for a given platform (excluding rack-level measurements). LinPACK power utilization is observed to go beyond processor TDP as it consist of a more complete power measurement. {The Marbl runs did not measure GPU-only power on Sierra.}}
  \label{fig:Measured power from Marbl}
\end{figure*}

Our starting point for the cross-platform analysis begins with understanding watts used for the simulation and the total runtime. Table~\ref{tab:Marbl_watts_used_cross_platform_energy} reports the rate at which energy is consumed. Figure~\ref{fig:Measured power from Marbl}
compares application power rates with other known values.

\input{tables/marbl_watts_used}

To compare across platforms, we introduce the concept of a CTS-1 \emph{work unit}, which we define as total number of quadrature points $\times$ cycles. Table~\ref{tab:Marbl_cross_platform_energy} presents Marbl's observed energy usage, throughput and comparisons to the CTS-1 platform. {For simulations on Sierra, it was necessary to run a much larger problem to measure the energy and power consumption}. After normalization, our findings show
that relative to CTS-1 performance, the GPU-based Sierra delivers a clear advantage in terms of improved energy efficiency (6.35x more efficient) and throughput (19.37x).  We also find that although Magma 
utilizes an energy quantity like the CTS-1 platform, it can deliver an almost three-fold throughput performance (2.78x). Lastly, we find that Astra can provide reduced energy usage (1.34x improvement to CTS-1)
for a 1.5x throughput improvement.

\input{tables/marbl_energy_study}

\subsection{Imp}

We designed our Imp simulations to run for 60 to 80 minutes on each
of three platforms -- CTS-1, Magma, and Sierra.  Like the Marbl cross platform study,
we compare energy and power usage across the smallest
number of nodes necessary to gather the data.

\begin{figure*}[tbp]
  \centering
  \includegraphics[width=.75\linewidth]{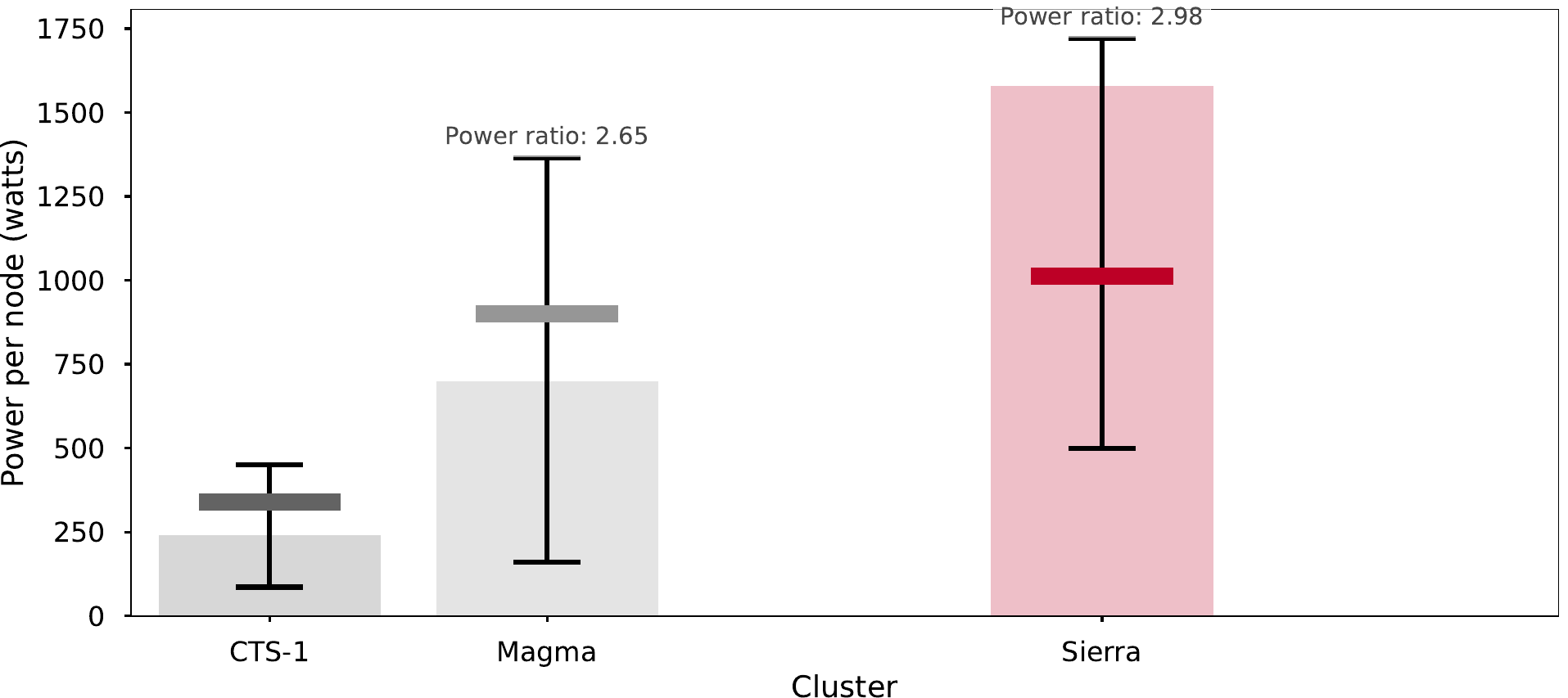}
  \caption{Imp power requirements (dark bar) relative to machine idle (lower whisker) and LinPACK (upper whisker) power rates. Power ratio refers to the necessary speed up to break even in terms of energy usage compared to the CTS-1 machine. Light colored bars corresponds to processor TDP for a given platform (excluding rack-level measurements). LinPACK power utilization is observed to go beyond processor TDP as it consist of a more complete power measurement. {Our Imp study did not include runs on the Astra platform, or GPU-only power measurements on Sierra.}}
  \label{fig:measured_power_imp}
\end{figure*}

Table~\ref{tab:imp_watts_used_cross_platform_energy} shows our cross platform comparison of the watts/node used by Imp. 
Unlike Ares and MARBL, Imp was not available to run on Astra and thus we omit results on Astra. 
Figure~\ref{fig:measured_power_imp} compares the application power rates with other known values.

To perform the comparison, we defined a work unit as the processing of $10^8$ photons.  We measured the
amount of work units completed, the seconds to solution, and the energy consumed.   After
normalization, we see that relative to CTS-1 performance, the Sierra system is 2.33x more efficient.
Also, Sierra provided 6.94x the throughput for the 2D hohlraum simulation.
Refer to Table~\ref{tab:imp_cross_platform_energy} for additional details of this study.

\input{tables/imp_watts_used}
\input{tables/imp_energy_study}

\section{Optimization impact on power and energy on Sierra}

\subsection{Ares}

\input{charts/ares_optimization}

To study the effects of optimization on power and energy on Sierra,
we tested two optimizations. The first was to run the code asynchronously as
long as possible, where the unoptimized version would synchronize after every
kernel. The second was to perform kernel fusion on the packing and unpacking kernels
that are used to compose the MPI communication buffers. These are very small kernels
that are kernel launch bound due to the small amount of data being moved per
kernel. This optimization is focused on improving strong scaling behavior.
The results of the tests running with and without all optimizations are
presented in Figure~\ref{figure:ares_optimization_chart} (and the data is presented in Table~\ref{tab:ares_optimizations_energy}).
As an additional baseline, the code was also run utilizing only the CPU cores on the platform. For these CPU-only
runs, we subtracted out the idle GPU power from Table~\ref{tab:idle_and_LinPACK_energy}
to have a fairer comparison.

Table~\ref{tab:ares_optimizations_energy} lists the data from the charts in Figure~\ref{figure:ares_optimization_chart}.

\input{tables/ares_optimization}

As noted in the cross-platform section, we see that the energy consumption
increases as we strong scale the problem under all problem configurations. We
also see that as we optimize the code, the strong scaling inefficiencies
lessen, which yields significantly lower energy usage at the highest node
counts.

When we look at the data across the optimizations, we see that the optimizations
all decrease the runtime of the problem and generally reduce the overall
energy consumption of the total run, except for the kernel fusion
optimization at 1 or 2 nodes, where they had little effect. There is a
trend in the data that as more optimizations are added, the power increases.
This is due to the optimizations eliminating gaps in between kernels, due to
kernel launch overheads, which keeps the GPU from idling, and thus does more
work in a shorter amount of time.
Although the power is increasing, the runtime of the problems
are decreasing faster, which leads to an overall gain in energy efficiency.

This data can also be used to discuss the energy breakeven point between using
the CPU and the GPU. Most simplistically, the ratio of the power used between
two runs can be used to determine how much faster the code needs to run to
have the same energy consumption. In comparing all the GPU runs to the CPU
runs, that ratio ranges between 1.15 and 1.42, depending on which node count
is used. Comparing the 16 node unoptimized run to the CPU only run confirms
this, as their energy usage is almost equal with only being 1.13x faster.
For every other comparison, the GPUs are clearly far more energy efficient
than using CPUs alone.

Another feature of the data to note is that although the vast majority of
the compute and memory bandwidth are being used by the GPU, the GPUs only
account for 35--45\% of the overall energy consumption of the node. This
reinforces that it is important to consider an entire architecture's node
design when looking at energy efficiency, rather than just the compute
units alone.

\subsection{Marbl}

To understand the impact on power and energy on Sierra, we aligned tracking power and energy usage on Sierra along with the GPU refactoring effort. At the start of this effort, Marbl strictly used MPI for parallelism; thus, porting to GPU platforms required significant refactoring. We begain tracking energy usage on Sierra shortly after reaching the node to node performance breakeven point where we had comparable runtimes when running the same problem on a CTS-1 node to a Sierra node utilizing the GPUs. We present our energy usage from May through October 2020. In general, we find that performance optimizations can affect power usage differently. Figure~\ref{figure:marbl_time_series}(a) illustrates that although not all optimizations led to increased power, there was a general downwards trend in total runtime. Total energy usage can be shown to decrease as the runtime performance of the code improves as shown in Figure~\ref{figure:marbl_time_series}(b).



\input{charts/marbl_power_timeseries}

\section{Conclusion}
In this work we have presented a methodology based on actual computing system energy measurements for understanding
power and energy consumption of production level scientific simulation codes. Using our methodology, we performed studies using the
Ares, Marbl, and Imp codes from Lawrence Livermore National Lab. We found that it is imperative to perform energy
measurements, as using a surrogate such as processor TDP may overestimate power usage when running scientific applications.

Additionally, we introduce the notion of the required speedup for an energy breakeven point between platforms.
Based on our experiments, we have found that Ares, Marbl, and Imp only require between 2-3x speedups on Sierra over a CTS-1 platform (node-to-node comparison)
to reach the breakeven point between platforms. In practice, however these codes achieve much greater speedups on a GPU-based system
compared to the CTS-1 platform thereby requiring less energy to run on the GPU platform. For these problems Sierra was able to provide an energy savings
of 2.33x for the Imp code, 6.35 for Marbl, and 1.6--5x for Ares over the CTS-1 platform.

\begin{figure*}[tbp]
  \centering
  \includegraphics[width=.75\linewidth]{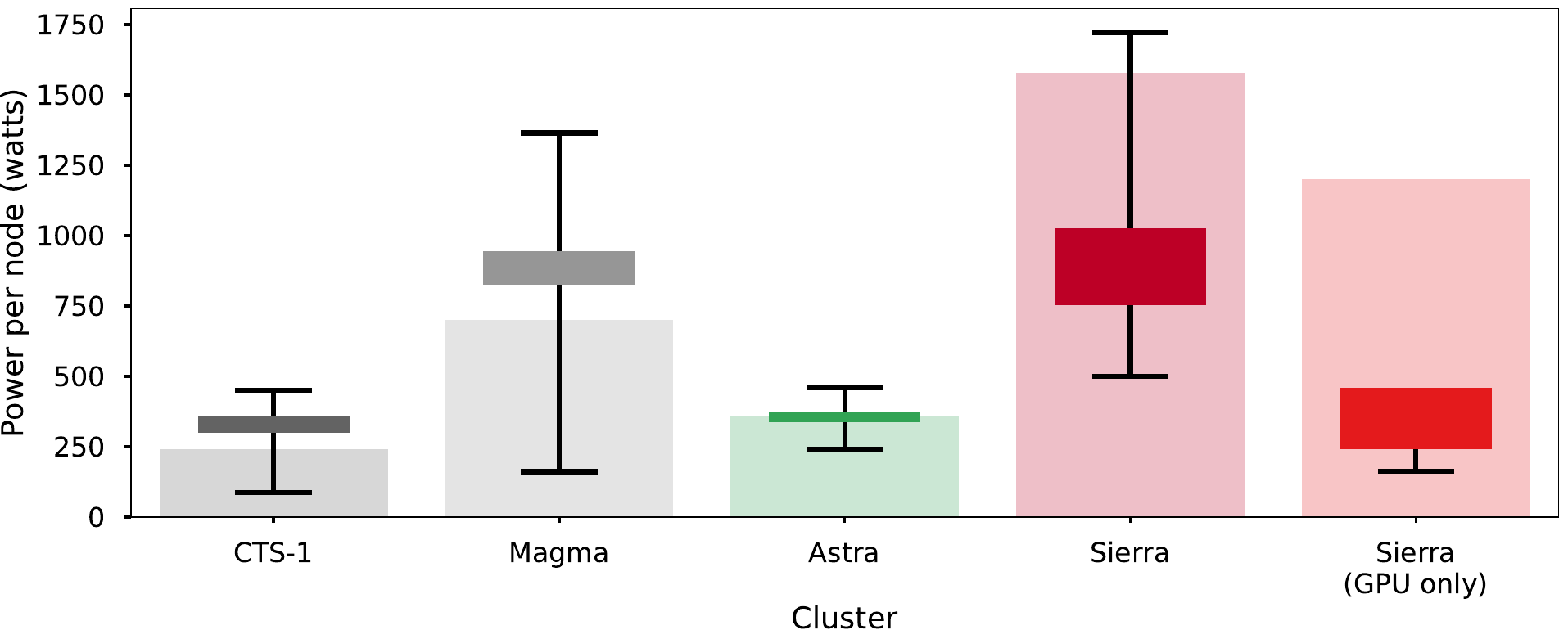}
  \caption{Application power requirements (dark bars) relative to machine idle (lower whisker) and LinPACK (upper whisker) power rates. Light colored bars corresponds to processor TDP for a given platform (excluding rack-level measurements). LinPACK power utilization is observed to go beyond processor TDP as it consist of a more complete power measurement.}
  \label{fig:Measured power from codes}
\end{figure*}

Our studies also suggest that faster execution time tends to result in improved energy efficiency, but often with greater power usage.
Figure~\ref{fig:Measured power from codes} illustrates the power ranges across for our three applications compared to the idle and LinPACK power rates. Finally, higher throughput correlates with higher energy efficiency.

\appendix

\section*{Acknowledgments} 
We would like to thank Teresa Bailey, Jim Silva, Rob Neely and Brian Pudliner for their support in this work. Additionally, we thank Barry Rountree, Stephanie Brink and the rest of the Variorum team for the useful discussions and feedback. This work performed under the auspices of the U.S. Department of Energy by Lawrence Livermore National Laboratory under Contract DE-AC52-07NA27344. LLNL-JRNL-865731.

\section{Ares optimization data}

Table~\ref{tab:ares_optimizations_energy} lists the data from the charts in Figure~\ref{figure:ares_optimization_chart}.

\bibliographystyle{unsrtnat}
\bibliography{main.bbl}  






\end{document}

%% file: tables/power_measurements.tex
\begin{table}[tbph]
  \centering
  \caption{Per-node power measurements on different platforms, measured in Watts}
  \label{tab:idle_and_LinPACK_energy}
  \begin{filecontents*}{data/power_measurements.csv}
platform          , tdp, idle, linpack
CTS-1             , 240, 86 , 450
Magma             , 700, 161 , 1365
Astra             , 360, 240 , 460
Sierra            , 1580, 500 , 1721
Sierra (GPUs only), 1200, 162 , nan
\end{filecontents*}
\pgfplotstabletypeset[
    multicolumn names,
    trim cells=true,
    col sep=comma,
    row sep=newline,
    every head row/.style={before row={\toprule},after row={\midrule}},
    every last row/.style={after row=\bottomrule},
    every row 4 column 3/.style={postproc cell content/.style={@cell content=\textit{not measured}}},    
    columns/platform/.style={column type=c, column name=Platform,string type},
    columns/tdp/.style={column type=r,column name=Processor TDP},
    columns/idle/.style={column type=r,column name=Power usage (idle)},
    columns/linpack/.style={column type=r,column name=Power usage (LinPACK)},
  ]
  {data/power_measurements.csv}
\end{table}

%% file: tables/ares_energy.tex
\begin{table}[tbph]
  \centering
  \caption{Energy and time measurements gathered from the various platforms with
           Ares, running the same problem, strong scaled across nodes.}
  \label{tab:ares_cross_platform_energy}

  \begin{filecontents*}{data/ares_energy.csv}
platform, nodes, duration, total_energy, avg_watts_per_node
CTS-1,    62,    28424,          167.56,             342.28
Magma,    48,    17094,          212.41,             931.96
Astra,    72,    20851,          148.37,             356.71
Astra,   144,    12705,          181.28,             355.70
Sierra,   5,     19755,           26.53,             967.23
Sierra,   10,    13106,           32.71,             898.52
Sierra,   20,     8967,           42.50,             853.23
Sierra,   40,     6979,           63.30,             816.30
Sierra,   80,     6240,          106.50,             768.05
\end{filecontents*}
\pgfplotstabletypeset[
    multicolumn names,
    trim cells=true,
    col sep=comma,
    row sep=newline,
    every head row/.style={before row={\toprule},after row={\midrule}},
    every last row/.style={after row=\bottomrule},
    columns/platform/.style={column type=l,column name={},string type},
    columns/nodes/.style={column type=r,column name=Nodes},
    columns/duration/.style={column type=r,column name=Duration (s)},
    columns/total_energy/.style={column type=r,column name=Total Energy (\kwh)},
    columns/avg_watts_per_node/.style={column type=r,column name=Avg. Watts Per Node},
    every row no 1/.append style={before row={\midrule}},
    every row no 2/.append style={before row={\midrule}},
    every row 3 column 0/.style={postproc cell content/.style={@cell content={}}},    
    every row no 4/.append style={before row={\midrule}},
    every row 5 column 0/.style={postproc cell content/.style={@cell content={}}},    
    every row 6 column 0/.style={postproc cell content/.style={@cell content={}}},    
    every row 7 column 0/.style={postproc cell content/.style={@cell content={}}},    
    every row 8 column 0/.style={postproc cell content/.style={@cell content={}}},    
    ]
  {data/ares_energy.csv}

  \end{table}

%% file: tables/marbl_watts_used.tex
\begin{table}[tbph]
  \centering
  \caption{Watts used in Marbl for 3D Triple point problem.}
  \label{tab:Marbl_watts_used_cross_platform_energy}
  \begin{filecontents*}{data/marbl_watts_used.csv}
  platform, nodes, duration, avg_watts_per_node, ratio_watts_to_CTS1
  CTS-1,      62,      2217,                314,                1.0    
  Magma,      48,      1024,                842,                2.68   
  Astra,      72,      1266,                351,                1.12   
  Sierra,     360,     1581,                958,                3.05   
\end{filecontents*}
\pgfplotstabletypeset[
    relativeCol/.style={postproc cell content/.append style={@cell content/.add={}{x},}},
    multicolumn names,
    trim cells=true,
    col sep=comma,
    row sep=newline,
    every head row/.style={before row={\toprule},after row={\midrule}},
    every last row/.style={after row=\bottomrule},
    every row 4 column 3/.style={postproc cell content/.style={@cell content=\textit{not measured}}},    
    columns/platform/.style={column type=c, column name=Platform,string type},
    columns/nodes/.style={column type=r,column name=Nodes},
    columns/duration/.style={column type=r,column name=Duration (s)},
    columns/avg_watts_per_node/.style={column type=r,column name=Avg. Watts Per Node},
    columns/ratio_watts_to_CTS1/.style={relativeCol,column type=r,column name=\parbox[c]{9em}{\centering Ratio of Watts used\\relative to CTS-1}},
  ]
  {data/marbl_watts_used.csv}
  \end{table}

%% file: tables/marbl_energy_study.tex
\begin{table}[tbph]
  \centering
  \caption{Cross platform study for Marbl}
  \label{tab:Marbl_cross_platform_energy}
  \begin{filecontents*}{data/marbl_energy_study.csv}
  platform , nodes, qpts,   cycles,  energy, throughput, rel_energy_efficiency , rel_throughput
  CTS-1    , 62   , 468e6,     500,  12.02 ,    1.92e10, 1,                      1
  Magma    , 48   , 462e6,     500,  11.49 ,    2.01e10, 1.04,                   2.78
  Astra    , 72   , 462e6,     500,  8.91  ,    2.59e10, 1.34,                   1.5
  Sierra   , 360  , 3.7e9,    5000,  151.5 ,    1.22e11, 6.35,                   19.37
\end{filecontents*}
\pgfplotstabletypeset[
    relativeCol/.style={postproc cell content/.append style={@cell content/.add={}{x},}},
    multicolumn names,
    trim cells=true,
    col sep=comma,
    row sep=newline,
    every head row/.style={before row={\toprule},after row={\midrule}},
    every last row/.style={after row=\bottomrule},
    every row 4 column 3/.style={postproc cell content/.style={@cell content=\textit{not measured}}},    
    columns/platform/.style={column type=c, column name=Platform,string type},
    columns/nodes/.style={column type=r,column name=Nodes},
    columns/qpts/.style={column type=r,column name=\parbox[c]{5em}{\centering Quadrature\\points}},
    columns/cycles/.style={column type=r,column name=Cycles},
    columns/duration/.style={column type=r,column name=Duration (s)},
    columns/energy/.style={column type=r,column name=\parbox[c]{4em}{\centering Total Energy\\(\kwh)}},
    columns/throughput/.style={column type=r,column name=\parbox[c]{6em}{\centering Throughput\\per \kwh}},
    columns/rel_energy_efficiency/.style={relativeCol,column type=r,column name=\parbox[c]{6em}{\centering Energy efficiency\\relative to CTS-1}},
    columns/rel_throughput/.style={relativeCol,column type=r,column name=\parbox[c]{6em}{\centering Throughput improvement\\relative to CTS-1}},
  ]
  {data/marbl_energy_study.csv}
  \end{table}

%% file: tables/imp_watts_used.tex
\begin{table}[tbph]
  \centering
  \caption{Watts used by Imp for 2D hohlraum simulation}
  \label{tab:imp_watts_used_cross_platform_energy}
  \begin{filecontents*}{data/imp_watts_used.csv}
platform, nodes,  work_units, duration, avg_watts_per_node, ratio_watts_to_CTS1
CTS-1,       62,         9.9,     3947,                340,        1.0
Magma,       48,        25.3,     4590,                901,        2.65
Sierra,      48,        51.2,     3798,               1012,        2.98
\end{filecontents*}
\pgfplotstabletypeset[
    relativeCol/.style={postproc cell content/.append style={@cell content/.add={}{x},}},
    multicolumn names,
    trim cells=true,
    col sep=comma,
    row sep=newline,
    every head row/.style={before row={\toprule},after row={\midrule}},
    every last row/.style={after row=\bottomrule},
    every row 4 column 3/.style={postproc cell content/.style={@cell content=\textit{not measured}}},    
    columns/platform/.style={column type=c, column name=Platform,string type},
    columns/nodes/.style={column type=r,column name=Nodes},
    columns/work_units/.style={column type=r,column name=Work units},
    columns/duration/.style={column type=r,column name=Duration (s)},
    columns/avg_watts_per_node/.style={column type=r,column name=Avg. Watts Per Node},
    columns/ratio_watts_to_CTS1/.style={relativeCol,column type=r,column name=\parbox[c]{9em}{\centering Ratio of Watts used\\relative to CTS-1}},
  ]
  {data/imp_watts_used.csv}
\end{table}

%% file: tables/imp_energy_study.tex
\begin{table}[tbph]
  \centering
  \caption{Cross platform study of 2D hohlraum simulation}
  \label{tab:imp_cross_platform_energy}
  \begin{filecontents*}{data/imp_energy_study.csv}
  platform , nodes, work_units,  duration,  energy, kwh_per_work_unit, rel_energy_efficiency , rel_throughput
  CTS-1,        62,        9.9,      3947,   23.12,              2.33,                    1,      1
  Magma,        48,        25.3,     4590,   55.14,              2.18,                  1.06,     2.81
  Sierra,       48,        51.2,     3798,   51.25,              1.00,                  2.33,     6.94
\end{filecontents*}
\pgfplotstabletypeset[
    relativeCol/.style={postproc cell content/.append style={@cell content/.add={}{x},}},
    multicolumn names,
    trim cells=true,
    col sep=comma,
    row sep=newline,
    every head row/.style={before row={\toprule},after row={\midrule}},
    every last row/.style={after row=\bottomrule},
    every row 4 column 3/.style={postproc cell content/.style={@cell content=\textit{not measured}}},    
    columns/platform/.style={column type=c, column name=Platform,string type},
    columns/nodes/.style={column type=r,column name=Nodes},
    columns/work_units/.style={column type=r,column name=\parbox[c]{3em}{\centering Work\\units}},
    columns/duration/.style={column type=r,column name=\parbox[c]{4em}{\centering Duration\\(s)}},
    columns/energy/.style={column type=r,column name=\parbox[c]{4em}{\centering Total Energy\\(\kwh)}},
    columns/kwh_per_work_unit/.style={column type=r,column name=\parbox[c]{6em}{\centering Throughput\\(\kwh\ per work unit)}},
    columns/rel_energy_efficiency/.style={relativeCol,column type=r,column name=\parbox[c]{6em}{\centering Energy efficiency\\relative to CTS-1}},
    columns/rel_throughput/.style={relativeCol,column type=r,column name=\parbox[c]{6em}{\centering Throughput improvement\\relative to CTS-1}},
  ]
  {data/imp_energy_study.csv}
\end{table}

%% file: charts/ares_optimization.tex
\newlength{\aresOptWidth}
\newlength{\aresOptHeight}
\setlength{\aresOptWidth}{.6\linewidth}
\setlength{\aresOptHeight}{2.5in}

\begin{figure}[hp]
  \centering
  \pgfplotstableread[col sep=comma]{data/ares_opt_total_energy.csv}\tableAresTotalEnergy
  \pgfplotstableread[col sep=comma]{data/ares_opt_gpu_energy.csv}\tableAresGPUEnergy
  \pgfplotstableread[col sep=comma]{data/ares_opt_duration.csv}\tableAresDuration
  \pgfplotstableread[col sep=comma]{data/ares_opt_power.csv}\tableAresPower

  \subfloat[Ares runtime]  {  
    \begin{tikzpicture}[trim axis right,trim axis left]
         \begin{axis}[
           legend pos=outer north east,
           xmode=log,
           log basis x=2,
           ymode=log,
           log basis y=2,
           width=\aresOptWidth,
           height=\aresOptHeight,
           xlabel={Nodes [log2]},
           ylabel={Runtime (s) [log2]},
         ]
         \addplot[color=Greys-3-3, very thick, mark=*] table [x=nodes, y=cpu_only] {\tableAresDuration}; \addlegendentry{CPU only}
         \addplot[color=Reds-7-4, very thick, mark=star] table [x=nodes, y=no_opt] {\tableAresDuration}; \addlegendentry{No opt.}
         \addplot[color=Reds-7-5, very thick, mark=triangle*] table [x=nodes, y=fusion_only] {\tableAresDuration}; \addlegendentry{Fusion only}
         \addplot[color=Reds-7-6, very thick, mark=diamond*] table [x=nodes, y=async_only] {\tableAresDuration}; \addlegendentry{Async. only}
         \addplot[color=Reds-7-7, very thick, mark=square*] table [x=nodes, y=full_opt] {\tableAresDuration}; \addlegendentry{Full opt.}
       \end{axis}      
     \end{tikzpicture}
   }  
  \\
  \subfloat[Ares energy usage]{
      \begin{tikzpicture}[trim axis right,trim axis left]
        \begin{axis}[
          legend pos=outer north east,
          xmode=log,
          log basis x=2,
          ymode=log,
          log basis y=2,
          width=\aresOptWidth,
          height=\aresOptHeight,
          xlabel={Nodes [log2]},
          ylabel={Energy usage (\kwh) [log2]},
          ymin=0,
        ]
        \addplot[color=Greys-3-3, very thick, mark=*] table [x=nodes, y expr=\thisrow{cpu_only}/3600000] {\tableAresTotalEnergy}; \addlegendentry{Total (CPU only)}
        \addplot[color=PuBu-5-2, very thick, mark=star] table [x=nodes, y expr=\thisrow{no_opt}/3600000] {\tableAresTotalEnergy}; \addlegendentry{Total (No opt.)}
        \addplot[color=PuBu-5-3, very thick, mark=triangle*] table [x=nodes, y expr=\thisrow{fusion_only}/3600000] {\tableAresTotalEnergy}; \addlegendentry{Total (Fusion only)}
        \addplot[color=PuBu-5-4, very thick, mark=diamond*] table [x=nodes, y expr=\thisrow{async_only}/3600000] {\tableAresTotalEnergy}; \addlegendentry{Total (Async. only)}
        \addplot[color=PuBu-5-5, very thick, mark=square*] table [x=nodes, y expr=\thisrow{full_opt}/3600000] {\tableAresTotalEnergy}; \addlegendentry{Total (Full opt.)}
        \addplot[color=PuRd-5-2, very thick, mark=star] table [x=nodes, y expr=\thisrow{no_opt}/3600000] {\tableAresGPUEnergy}; \addlegendentry{GPU (No opt.)}
        \addplot[color=PuRd-5-3, very thick, mark=triangle*] table [x=nodes, y expr=\thisrow{fusion_only}/3600000] {\tableAresGPUEnergy}; \addlegendentry{GPU (Fusion only)}
        \addplot[color=PuRd-5-4, very thick, mark=diamond*] table [x=nodes, y expr=\thisrow{async_only}/3600000] {\tableAresGPUEnergy}; \addlegendentry{GPU (Async. only)}
        \addplot[color=PuRd-5-5, very thick, mark=square*] table [x=nodes, y expr=\thisrow{full_opt}/3600000] {\tableAresGPUEnergy}; \addlegendentry{GPU (Full opt.)}
        \addplot[color=Greys-3-3, very thick, mark=*] table [x=nodes, y expr=\thisrow{cpu_only}/3600000] {\tableAresTotalEnergy};
      \end{axis}           
    \end{tikzpicture}
  }
  \\
  \subfloat[Ares average power per node]  {  
   \begin{tikzpicture}[trim axis right,trim axis left]
        \begin{axis}[
          legend pos=outer north east,
          xmode=log,
          log basis x=2,
          width=\aresOptWidth,
          height=\aresOptHeight,
          xlabel={Nodes [log2]},
          ylabel=Average Watts per node,
          ymin=500,
          axis y discontinuity=crunch,
        ]
        \addplot[color=Greys-3-3, very thick, mark=*] table [x=nodes, y=cpu_only] {\tableAresPower}; \addlegendentry{CPU only}
        \addplot[color=Blues-7-4, very thick, mark=star] table [x=nodes, y=no_opt] {\tableAresPower}; \addlegendentry{No opt.}
        \addplot[color=Blues-7-5, very thick, mark=triangle*] table [x=nodes, y=fusion_only] {\tableAresPower}; \addlegendentry{Fusion only}
        \addplot[color=Blues-7-6, very thick, mark=diamond*] table [x=nodes, y=async_only] {\tableAresPower}; \addlegendentry{Async. only}
        \addplot[color=Blues-7-7, very thick, mark=square*] table [x=nodes, y=full_opt] {\tableAresPower}; \addlegendentry{Full opt.}
      \end{axis}      
    \end{tikzpicture}
  }
  \caption{Comparing the affects of several optimizations on Ares runtime (a), energy (b) and average power (c) in a strong scaling study.
  }
  \label{figure:ares_optimization_chart}
\end{figure}

%% file: tables/ares_optimization.tex
\begin{table}[p]
  \centering
  
  \begin{tabular}{|l|c|c|c|c|c|}
  \hline
            & \multicolumn{5}{c|}{\textbf{Sierra Energy with Optimizations (\kwh)}} \\ \cline{2-6}
            & Unopt   & Async opt & Fusion opt  & Full opt & CPU only \\ \hline
  1  node   & 1.80    & 1.77      & 1.92        & 1.80     & X        \\ \hline
  2  nodes  & 2.14    & 2.00      & 2.15        & 1.98     & 6.92     \\ \hline
  4  nodes  & 2.86    & 2.41      & 2.72        & 2.45     & 7.26     \\ \hline
  8  nodes  & 4.14    & 3.35      & 3.63        & 3.23     & 7.17     \\ \hline
  16 nodes  & 7.70    & 5.50      & 6.22        & 4.76     & 7.50     \\ \hline
  \end{tabular}
  
  \vspace{7mm}
  
  \begin{tabular}{|l|c|c|c|c|c|}
  \hline
            & \multicolumn{5}{c|}{\textbf{Sierra GPU Energy with Optimizations (\kwh)}} \\ \cline{2-6}
            & Unopt   & Async opt & Fusion opt  & Full opt & CPU only \\ \hline
  1  node   & 0.84    & 0.81      & 0.86        & 0.81     & X        \\ \hline
  2  nodes  & 0.92    & 0.87      & 0.91        & 0.85     & X        \\ \hline
  4  nodes  & 1.18    & 1.02      & 1.10        & 1.03     & X        \\ \hline
  8  nodes  & 1.45    & 1.30      & 1.32        & 1.28     & X        \\ \hline
  16 nodes  & 2.57    & 1.90      & 2.11        & 1.70     & X        \\ \hline
  \end{tabular}
  
  \vspace{7mm}
  
  \begin{tabular}{|l|c|c|c|c|c|}
  \hline
            & \multicolumn{5}{c|}{\textbf{Sierra Average Watts per Node}} \\ \cline{2-6}
            & Unopt   & Async opt & Fusion opt  & Full opt & CPU only \\ \hline
  1  node   & 756     & 779       & 817         & 793      & X        \\ \hline
  2  nodes  & 777     & 795       & 801         & 795      & 591      \\ \hline
  4  nodes  & 768     & 758       & 773         & 782      & 595      \\ \hline
  8  nodes  & 685     & 736       & 683         & 750      & 587      \\ \hline
  16 nodes  & 660     & 684       & 666         & 675      & 572      \\ \hline
  \end{tabular}
  
  \vspace{7mm}
  
  \begin{tabular}{|l|c|c|c|c|c|}
  \hline
            & \multicolumn{5}{c|}{\textbf{Sierra Duration (seconds)}} \\ \cline{2-6}
            & Unopt   & Async opt & Fusion opt  & Full opt & CPU only \\ \hline
  1  node   & 8572    & 8179      & 8463        & 8162     & X        \\ \hline
  2  nodes  & 4977    & 4523      & 4839        & 4491     & 21084    \\ \hline
  4  nodes  & 3349    & 2859      & 3172        & 2819     & 10977    \\ \hline
  8  nodes  & 2720    & 2051      & 2388        & 1941     & 5491     \\ \hline
  16 nodes  & 2625    & 1807      & 2104        & 1587     & 2968     \\ \hline
  \end{tabular}
  
  \caption{Ares optimizations }
  \label{tab:ares_optimizations_energy}
  \end{table}

%% file: charts/marbl_power_timeseries.tex
\newlength{\figWidth}
\newlength{\figHeight}
\setlength{\figWidth}{.425\linewidth}
\setlength{\figHeight}{2.5in}

\begin{figure}[t]
  \centering
  \pgfplotstableread[col sep=comma]{data/marbl_power_timeseries.csv}\tableMarblTimeseries
  \subfloat[Marbl power and runtime]{
     \begin{tikzpicture}
        \begin{axis}[
          width=\figWidth,
          height=\figHeight,
          separate axis lines,
          y axis line style={darkerBlue},
          y label style={darkerBlue},
          y tick style={darkerBlue},
          y tick label style={darkerBlue},
          xlabel=Date,
          ylabel=Watts,
          date coordinates in=x,
          xticklabel style={rotate=45, anchor=east},
          xtick={2020-05-01, 2020-06-01, 2020-07-01, 2020-08-01, 2020-09-01, 2020-10-01},
          xticklabel=\year-\month,
          ymin=0,
          ymax=1000,
          axis y line*=left,
        ]
        \addplot[solid,darkerBlue,very thick,dashed, mark=X]   table [x=date, y=avg_watts_per_node] {\tableMarblTimeseries}; \label{chart:marbl_watts_per_node}
      \end{axis}      
      \begin{axis}[
          width=\figWidth,
          height=\figHeight,
          separate axis lines,
          y axis line style={red},
          y label style={red},
          y tick style={red},
          y tick label style={red},
          minor y tick num=1,
          ytick={0,200,400,600,800,1000,1200},
          xlabel=Date,
          ylabel=Runtime (s),
          date coordinates in=x,
          xticklabel style={rotate=45, anchor=east},
          xtick={2020-05-01, 2020-06-01, 2020-07-01, 2020-08-01, 2020-09-01, 2020-10-01},
          xticklabel=\year-\month,
          axis y line*=right,
          axis x line=none,
          ymin=0,
          ymax=1200,
        ]
        \addlegendimage{/pgfplots/refstyle=chart:marbl_watts_per_node}\addlegendentry{Avg. Watts per node}
        \addplot[solid,red,very thick, mark=*]   table [x=date, y=run_time] {\tableMarblTimeseries};\addlegendentry{Simulation run time}
      \end{axis}      
    \end{tikzpicture}
  }
  \subfloat[Marbl energy usage]{
      \begin{tikzpicture}
        \begin{axis}[
          width=\figWidth,
          height=\figHeight,
          xlabel=Date,
          ylabel=Energy usage (\kwh),
          date coordinates in=x,
          xticklabel style={rotate=45, anchor=east},
          xtick={2020-05-01, 2020-06-01, 2020-07-01, 2020-08-01, 2020-09-01, 2020-10-01},
          xticklabel=\year-\month,
          ymin=0,
          yticklabel style={/pgf/number format/fixed},
        ]
        \addplot[color=PuBu-5-4, very thick,mark=*] table [x=date, y expr=\thisrow{total_energy}/3600000] {\tableMarblTimeseries}; \addlegendentry{Total energy}
        \addplot[color=PuRd-5-4, very thick, mark=*] table [x=date, y expr=\thisrow{gpu_energy}/3600000] {\tableMarblTimeseries}; \addlegendentry{GPU energy}
      \end{axis}           
    \end{tikzpicture}
  }
  \caption{Tracked runtime, power and energy usage for Marbl's 3D Shaped Charge problem during an active phase of its GPU refactoring.
    Data was captured on a single node of Sierra (4 GPUs) between May and October 2020.
    (a) Simulation average power usage (dashed blue line, using left axis) and runtime (red solid line, using the right axis). 
    (b) Energy usage across the entire node (dark blue) and restricted to the GPUs (magenta).  
  }
  \label{figure:marbl_time_series}
\end{figure}

%% file: main.bbl
\begin{thebibliography}{23}
\providecommand{\natexlab}[1]{#1}
\providecommand{\url}[1]{\texttt{#1}}
\expandafter\ifx\csname urlstyle\endcsname\relax
  \providecommand{\doi}[1]{doi: #1}\else
  \providecommand{\doi}{doi: \begingroup \urlstyle{rm}\Url}\fi

\bibitem[{Green 500}(2024)]{top500green}
{Green 500}.
\newblock https://www.top500.org/lists/green500, 2024.
\newblock URL \url{https://www.top500.org/lists/green500/}.

\bibitem[Franko et~al.(2015)Franko, Fisher, Lin, and Bova]{franko2015cfd}
Kenneth Franko, Travis~C Fisher, Paul Lin, and Steven~W Bova.
\newblock {CFD} for next generation hardware: Experiences with proxy
  applications.
\newblock In \emph{22nd AIAA Computational Fluid Dynamics Conference}, page
  3053, 2015.

\bibitem[Lannelongue et~al.(2021)Lannelongue, Grealey, and
  Inouye]{lannelongue2021green}
Lo{\"\i}c Lannelongue, Jason Grealey, and Michael Inouye.
\newblock Green algorithms: quantifying the carbon footprint of computation.
\newblock \emph{Advanced science}, 8\penalty0 (12):\penalty0 2100707, 2021.

\bibitem[McFadden et~al.(2019)McFadden, Shoga, Brink, Rountree, Patki,
  Cantalupo, Guttman, Geltz, and Allen]{mcfadden2019msr}
Martin~J McFadden, Kathleen~S Shoga, Stephanie Brink, Barry~L Rountree, Tapasya
  Patki, Christopher Cantalupo, Diana Guttman, Brad Geltz, and Ben Allen.
\newblock msr-safe.
\newblock Technical report, Lawrence Livermore National Laboratory (LLNL),
  Livermore, CA (United States), 2019.

\bibitem[Brink et~al.(2024)Brink, Marathe, Patki, and Rountree]{Variorum}
Stephanie Brink, Aniruddha Marathe, Tapasya Patki, and Barry Rountree.
\newblock Variorum.
\newblock \url{https://github.com/LLNL/variorum}, 2024.

\bibitem[Enos et~al.(2010)Enos, Steffen, Fullop, Showerman, Shi, Esler,
  Kindratenko, Stone, and Phillips]{enos2010quantifying}
Jeremy Enos, Craig Steffen, Joshi Fullop, Michael Showerman, Guochun Shi,
  Kenneth Esler, Volodymyr Kindratenko, John~E Stone, and James~C Phillips.
\newblock Quantifying the impact of {GPU}s on performance and energy efficiency
  in {HPC} clusters.
\newblock In \emph{International Conference on Green Computing}, pages
  317--324. IEEE, 2010.

\bibitem[Patel et~al.(2020)Patel, Wagenh{\"a}user, Eibel, H{\"o}nig, Zeiser,
  and Tiwari]{patel2020does}
Tirthak Patel, Adam Wagenh{\"a}user, Christopher Eibel, Timo H{\"o}nig, Thomas
  Zeiser, and Devesh Tiwari.
\newblock What does power consumption behavior of {HPC} jobs reveal?:
  Demystifying, quantifying, and predicting power consumption characteristics.
\newblock In \emph{2020 IEEE International Parallel and Distributed Processing
  Symposium (IPDPS)}, pages 799--809. IEEE, 2020.

\bibitem[Kamil et~al.(2008)Kamil, Shalf, and Strohmaier]{kamil2008power}
Shoaib Kamil, John Shalf, and Erich Strohmaier.
\newblock Power efficiency in high performance computing.
\newblock In \emph{2008 IEEE International Symposium on Parallel and
  Distributed Processing}, pages 1--8. IEEE, 2008.

\bibitem[Horwitz(2024)]{horwitz2024estimating}
JAK Horwitz.
\newblock Estimating the carbon footprint of {Computational Fluid Dynamics}.
\newblock \emph{Physics of Fluids}, 36\penalty0 (4), 2024.

\bibitem[{Top 500}(2024)]{top500}
{Top 500}.
\newblock https://www.top500.org/lists/top500, 2024.
\newblock URL \url{https://www.top500.org/lists/top500}.

\bibitem[Grant et~al.(2023)Grant, Hammond, Laros~III, Levenhagen, Olivier,
  Pedretti, Ward, and Younge]{grant2023enabling}
Ryan~E Grant, Simon~D Hammond, James~H Laros~III, Michael Levenhagen, Stephen~L
  Olivier, Kevin Pedretti, Lee Ward, and Andrew~J Younge.
\newblock Enabling power measurement and control on {Astra}: The first
  petascale {ARM} supercomputer.
\newblock \emph{Concurrency and Computation: Practice and Experience},
  35\penalty0 (15):\penalty0 e7303, 2023.

\bibitem[Grant et~al.(2016)Grant, Levenhagen, Olivier, DeBonis, Pedretti, and
  Laros~III]{grant2016standardizing}
Ryan~E Grant, Michael Levenhagen, Stephen~L Olivier, David DeBonis, Kevin~T
  Pedretti, and James~H Laros~III.
\newblock Standardizing power monitoring and control at exascale.
\newblock \emph{Computer}, 49\penalty0 (10):\penalty0 38--46, 2016.

\bibitem[Beckingsale et~al.(2019{\natexlab{a}})Beckingsale, Burmark, Hornung,
  Jones, Killian, Kunen, Pearce, Robinson, Ryujin, and
  Scogland]{beckingsale2019raja}
David~A Beckingsale, Jason Burmark, Rich Hornung, Holger Jones, William
  Killian, Adam~J Kunen, Olga Pearce, Peter Robinson, Brian~S Ryujin, and
  Thomas~RW Scogland.
\newblock {RAJA}: Portable performance for large-scale scientific applications.
\newblock In \emph{2019 IEEE/ACM international workshop on performance,
  portability and productivity in HPC (p3hpc)}, pages 71--81. IEEE,
  2019{\natexlab{a}}.

\bibitem[Beckingsale et~al.(2019{\natexlab{b}})Beckingsale, Mcfadden, Dahm,
  Pankajakshan, and Hornung]{beckingsale2019umpire}
David~A Beckingsale, Marty~J Mcfadden, Johann~PS Dahm, Ramesh Pankajakshan, and
  Richard~D Hornung.
\newblock Umpire: Application-focused management and coordination of complex
  hierarchical memory.
\newblock \emph{IBM Journal of Research and Development}, 64\penalty0
  (3/4):\penalty0 00--1, 2019{\natexlab{b}}.

\bibitem[Bender et~al.(2021)Bender, Schilling, Raman, Managan, Olson, Copeland,
  Ellison, Erskine, Huntington, Morgan, and et~al.]{bender_ares_2021}
Jason~D. Bender, Oleg Schilling, Kumar~S. Raman, Robert~A. Managan, Britton~J.
  Olson, Sean~R. Copeland, C.~Leland Ellison, David~J. Erskine, Channing~M.
  Huntington, Brandon~E. Morgan, and et~al.
\newblock Simulation and flow physics of a shocked and reshocked
  high-energy-density mixing layer.
\newblock \emph{Journal of Fluid Mechanics}, 915:\penalty0 A84, 2021.
\newblock \doi{10.1017/jfm.2020.1122}.

\bibitem[Anderson et~al.(2020)Anderson, Black, Blakeley, Bleile, Camier,
  Ciurej, Cook, Dobrev, Elliott, Grondalski, Harrison, Hornung, Kolev,
  Legendre, Liu, Nissen, Olson, Osawe, Papadimitriou, Pearce, Pember, Skinner,
  Stevens, Stitt, Taylor, Tomov, Rieben, Vargas, Weiss, White, and
  Busby]{anderson2020multiphysics}
R~Anderson, A.~Black, B.~Blakeley, R.~Bleile, J.-S. Camier, J.~Ciurej, A.~Cook,
  V.~Dobrev, N.~Elliott, J.~Grondalski, C.~Harrison, R.~Hornung, Tz. Kolev,
  M.~Legendre, W.~Liu, W.~Nissen, B.~Olson, M.~Osawe, G.~Papadimitriou,
  O.~Pearce, R.~Pember, A.~Skinner, D.~Stevens, T.~Stitt, L.~Taylor, V.~Tomov,
  R.~Rieben, A.~Vargas, K.~Weiss, D.~White, and L.~Busby.
\newblock The {Multiphysics} on {Advanced} {Platforms} {Project}.
\newblock Technical report, Technical Report LLNL-TR-815869, LLNL, 2020.

\bibitem[Capps et~al.(2017--2024)Capps, Carson, Corbett, Elliott, Essman,
  Gunney, Han, Harrison, Hornung, Larsen, Moody, Pauli, Settgast, Taylor,
  Weiss, White, Whitlock, Yang, and Zagaris]{Axom}
A.~Capps, R.~Carson, B.~Corbett, N.~Elliott, J.~Essman, B.~Gunney, B.~Han,
  C.~Harrison, R.~Hornung, M.~Larsen, A.~Moody, E.~Pauli, R.~Settgast,
  L.~Taylor, K.~Weiss, C.~White, B.~Whitlock, M.~Yang, and G.~Zagaris.
\newblock Axom: {CS} infrastructure components for {HPC} applications,
  2017--2024.
\newblock URL \url{https://github.com/llnl/axom}.

\bibitem[Anderson et~al.(2021)Anderson, Andrej, Barker, Bramwell, Camier,
  Cerveny, Dobrev, Dudouit, Fisher, Kolev, Pazner, Stowell, Tomov, Akkerman,
  Dahm, Medina, and Zampini]{mfem}
R.~Anderson, J.~Andrej, A.~Barker, J.~Bramwell, J.-S. Camier, J.~Cerveny,
  V.~Dobrev, Y.~Dudouit, A.~Fisher, Tz. Kolev, W.~Pazner, M.~Stowell, V.~Tomov,
  I.~Akkerman, J.~Dahm, D.~Medina, and S.~Zampini.
\newblock {MFEM}: A modular finite element methods library.
\newblock \emph{Computers \& Mathematics with Applications}, 81:\penalty0
  42--74, 2021.
\newblock \doi{10.1016/j.camwa.2020.06.009}.

\bibitem[Vargas et~al.(2022)Vargas, Stitt, Weiss, Tomov, Camier, Kolev, and
  Rieben]{Vargas2022_ijhpca}
Arturo Vargas, Thomas~M. Stitt, Kenneth Weiss, Vladimir~Z. Tomov, Jean-Sylvain
  Camier, Tzanio Kolev, and Robert~N. Rieben.
\newblock Matrix-free approaches for {GPU} acceleration of a high-order finite
  element hydrodynamics application using {MFEM}, {Umpire}, and {RAJA}.
\newblock \emph{The International Journal of High Performance Computing
  Applications}, 36\penalty0 (4):\penalty0 492--509, May 2022.
\newblock \doi{10.1177/10943420221100262}.

\bibitem[Stitt et~al.(2024)Stitt, Belcher, Campos, Kolev, Mocz, Rieben,
  Skinner, Tomov, Vargas, and Weiss]{Stitt2024_jfe}
Thomas Stitt, Kristi Belcher, Alejandro Campos, Tzanio Kolev, Philip Mocz,
  Robert~N. Rieben, Aaron Skinner, Vladimir Tomov, Arturo Vargas, and Kenneth
  Weiss.
\newblock Performance portable {Graphics Processing Unit} acceleration of a
  high-order finite element multiphysics application.
\newblock \emph{Journal of Fluids Engineering}, 146\penalty0 (4):\penalty0
  041102, 02 2024.
\newblock ISSN 0098-2202.
\newblock \doi{10.1115/1.4064493}.

\bibitem[Brantley et~al.(2019)Brantley, Gentile, Lambert, McKinley, O'Brien,
  and A.]{Brantley2019}
P.~S. Brantley, N.~A. Gentile, M.~A. Lambert, M.~S. McKinley, M.~J. O'Brien,
  and Walsh~J. A.
\newblock A new implicit {M}onte {C}arlo thermal photon transport capability
  developed using shared {Monte} {Carlo} infrastructure.
\newblock \emph{The International Conference on Mathematics and Computational
  Methods Applied to Nuclear Science and Engineering, Portland, Oregon, August
  25-29, 2019}, 2019.

\bibitem[Fleck~Jr and Cummings~Jr(1971)]{fleck1971implicit}
JA~Fleck~Jr and JD~Cummings~Jr.
\newblock An implicit {Monte} {Carlo} scheme for calculating time and frequency
  dependent nonlinear radiation transport.
\newblock \emph{Journal of Computational Physics}, 8\penalty0 (3):\penalty0
  313--342, 1971.

\bibitem[Yee et~al.(2021)Yee, Olivier, Southworth, Holec, and Haut]{yee2021new}
Ben~C Yee, Samuel~S Olivier, Ben~S Southworth, Milan Holec, and Terry~S Haut.
\newblock A new scheme for solving high-order {DG} discretizations of thermal
  radiative transfer using the variable {Eddington} factor method.
\newblock \emph{arXiv preprint arXiv:2104.07826}, 2021.

\end{thebibliography}
